# Efficient Tree Solver for Hines Matrices on the GPU
## using fine grained parallelization and basic work balancing

Felix Huber
Institute of Applied Analysis
and Numerical Simulation
University of Stuttgart
Germany
f.huber@mathematik.uni-stuttgart.de

**Abstract** *The human brain consists of a large number of interconnected neurons communicating via exchange of electrical spikes. Simulations play an important role in better understanding electrical activity in the brain and offers a way to to compare measured data to simulated data such that experimental data can be interpreted better. A key component in such simulations is an efficient solver for the Hines matrices used in computing inter-neuron signal propagation. In order to achieve high performance simulations, it is crucial to have an efficient solver algorithm. In this report we explain a new parallel GPU solver for these matrices which offers fine grained parallelization and allows for work balancing during the simulation setup.*

## 1 Modeling the Neuron

While neurons can strongly differ in their size and structure, they all consist of three parts: the *dendrites* receive signals from other neurons in a tree structure and forwards them to the *soma*. If the signal that reaches the soma rises over a certain threshold, the soma emits a new signal over the *axon* which leads to the ends of dendrites of other neurons. A simplified structure of a neuron is given in Figure 1.1.

In Arbor [1] the axons are modeled by a time delay, since the shape of the spike does not change significantly while traveling over the axon. When a soma emits a new spike, arbor does not explicitly simulate the transport over the axon, but instead inserts the spike at the end of a dendrite of another neuron with a delayed time of arrival. Since the dendrite collects spikes from many different neurons and overlays the signals on it's way to the soma, we can't model them by a simple time delay. Instead we have to explicitly compute how the signals interact in the dendrite over time.

The model used is based on the Hodgkin and Huxley model [6], which describes the transport process of the action potential on the cell membrane. Since each dendrite is very thin compared to its length, we can model each branch by a one dimensional, radially symmetric electric cable. This leads to the parabolic nonlinear *cable equation* [4] for the voltage $V(x)$ along a branch

$$c_m \frac{\partial V}{\partial t} = \frac{1}{2\pi a r_L} \frac{\partial}{\partial x}\left(a^2 \frac{\partial V}{\partial x}\right) + I$$

with the membrane capacity $c_m$, the cable radius $a(x)$, the intracellular resistivity $r_L$ and the current $I(V)$ from electrodes, ion channels and synapses. This PDE is nonlinear since the the membrane current in $I(V)$ depends on the voltage.

Splitting each branch into compartments and using a finite volume scheme with an implicit Euler method for the integration over the time step, we get the discrete linear system

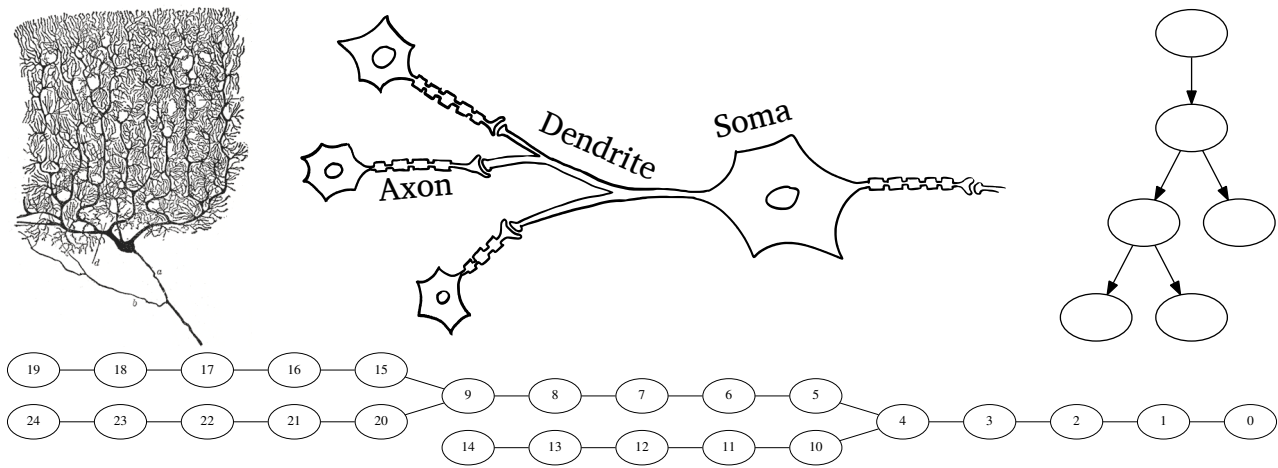

Figure 1.1: Summary of the different modeling steps. In the top center we see a model of a neuron consisting of the soma, dendrites and axon. Real neurons can get very complex as shown in the example on the left [3]. If we subdivide each branch in the dendrite tree into five compartments we get the tree below which shows how the compartments are connected. The numbering of the compartment ensures that we get tridiagonal submatrices. The tree on the top right is the dependency tree of the tasks during a backward substitution. Each node corresponds to a branch in the dendrite tree where we have to perform a backwards substitution on a tridiagonal submatrix.

of equations [1, documentation]

$$\left(\frac{\sigma_i c_{m,i}}{\Delta t} + \sum_{j \in \mathcal{N}_i} a_{ij}\right) V_i^{k+1} - \sum_{j \in \mathcal{N}_i} a_{ij} V_j^{k+1} = \frac{\sigma_i c_{m,i}}{\Delta t} V_i^k + \sigma_i I$$

where we avoid the possible nonlinearity of $I(V)$ by evaluating it at the old time step $k$ instead of the new time step $k+1$. Here, $\mathcal{N}_i$ is the set of all neighboring compartments and for the coefficients $a_{ij} = a_{ji}$. Since all $\sigma_i$ and $c_{m,i}$ are positive the resulting matrix is symmetric, strictly diagonally dominant and strictly positive definite.

Hence, each dendrite branch in the tree structure of the dendrite leads to a tridiagonal matrix. When we compute the complete matrix for the entire dendrite tree we want to ensure that we keep the structure of the tridiagonal submatrices. However, these submatrices are not independent from each other. Because the ends of the branches are connected at the branch points of the dendritic tree, the combined matrix also contains entries that couple the tridiagonal submatrices. This system of equations has to be solved in every time step of the simulation.

While tridiagonal matrices can be solved easily with the Thomas algorithm [5], we have to modify this algorithm to keep track of the parent compartment of each unknown. Instead of assuming that the parent of variable $i$ is always $i-1$, we have to lookup the parent in each elimination steps. This algorithm is called the Hines algorithm [7].

## 2 HINES SOLVER ALGORITHM

The key idea for solving these matrices is to use a good numbering scheme for the unknown variables, so that an adapted version of Thomas algorithms can be used to solve the problem. An important observation is that the tree structure of the dendrites resembles the dependencies between the submatrices. If we number each dendrite branch breadth first from the root of the tree we can ensure that during the backward substitution the rows only depend on the solutions of submatrices with larger indices. This resulting matrix structure is called a *Hines matrix*. An example for a such a numbering and the resulting matrix is given in Figure 1.1 and Figure 2.1. As with the Thomas algorithm we can now solve the matrix by performing a backward substitution followed by a forward substitution. The only difference from the Thomas algorithm is that at the end of each backward substitution on a submatrix we don't just eliminate a matrix entry in the previous row, but instead we eliminate one coupling entry in the parent branch. Similarly at the beginning of each forward substitution eliminate the entry for the dependency on the parent branch. This means that the solver needs an additional array describing the parent compartment for each branch. This algorithm is called the *Hines solver* [7]. Since we still solve the matrix with exactly one backward and one forward substitution, the complexity of the solver is still $\mathcal{O}(N)$ where $N$ is the number of unknowns. In the following we show how the Thomas algorithms can be adapted into the Hines algorithm by using an array describing the parent relations. In the algorithms, $b$ denotes the right hand side of the matrix, $d$ the diagonal entries and $u$ the upper diagonal entries. For the Hines algorithm $u$ also stores the coupling entries to the parent of a compartment. Additionally, the Hines algorithm takes $p$ as input, which stores the parent compartment for each compartment. If $p[i] = i - 1$ holds, then Hines algorithm is equivalent to Thomas algorithm.

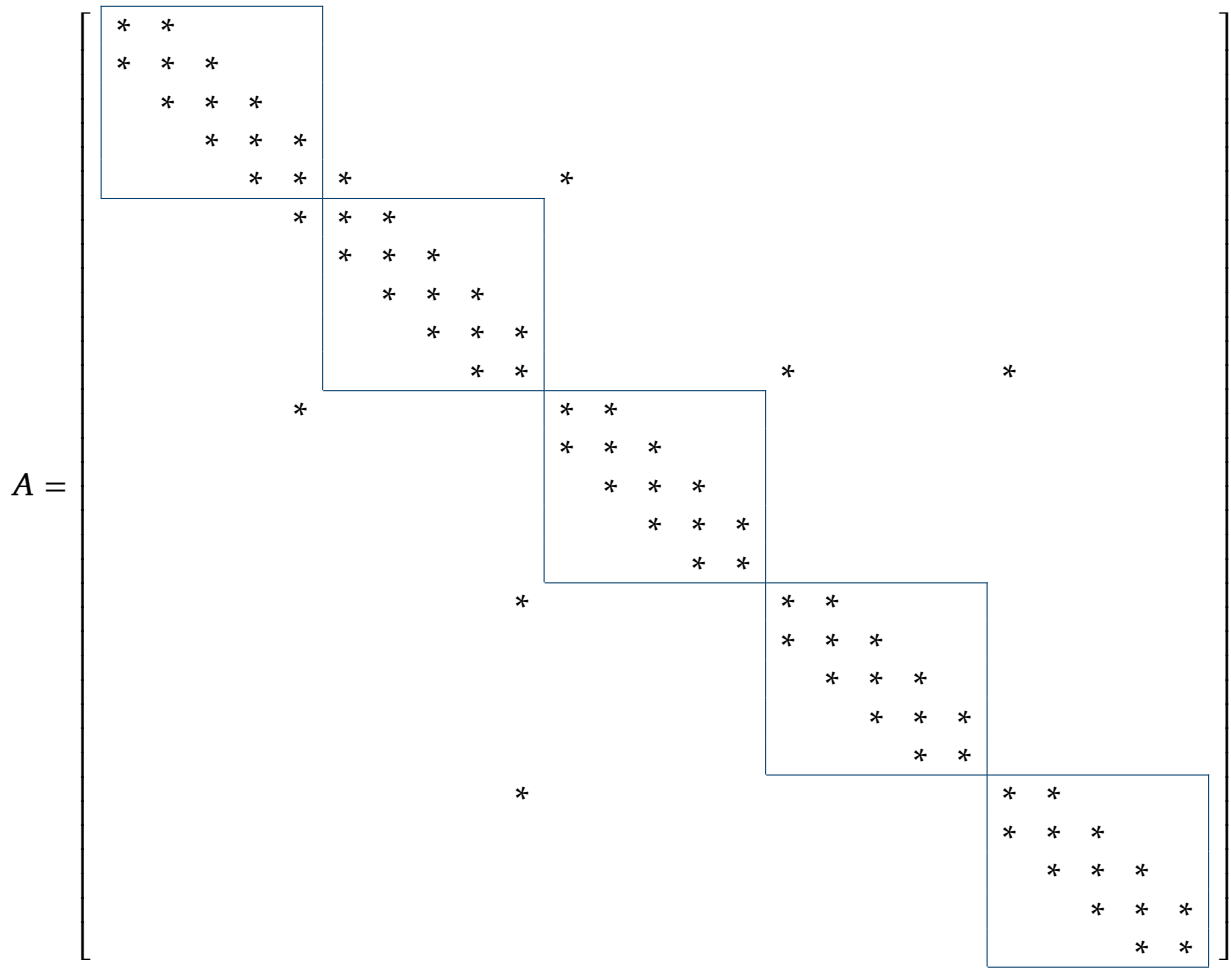

Figure 2.1: An example of a Hines matrix for the dendritic tree in Figure 1.1, where each branch is subdivided into five compartments. Each tridiagonal matrix block corresponds to one branch in the dendritic tree. The block in the upper left corresponds to the root branch with the root $A_{0,0}$. The entries outside the tridiagonal structure describe the coupling in the tree structure.

LISTING 1: THOMAS ALGORITHM FOR SYMMETRIC MATRICES

```
def thomas(d, u, b):
    # backward substitution
    for i in N-1,...,1:
        f = u[i-1] / d[i]
        d[i-1] = d[i-1] - f * u[i-1]
        b[i-1] = b[i-1] - f * b[i]

    # solve root
    b[0] = b[0] / d[0]

    # forward substitution
    for i in 1,...,N-1:
        b[i] = (b[i] - u[i] * b[i-1]) / d[i]

    return b
```

LISTING 2: HINES ALGORITHM FOR SYMMETRIC MATRICES

```
def hines(d, u, b, p):
    # backward substitution
    for i in N-1,...,1:
        f = u[p[i]] / d[i]
        d[p[i]] = d[p[i]] - f * u[p[i]]
        b[p[i]] = b[p[i]] - f * b[i]

    # solve root
    b[0] = b[0] / d[0]

    # forward substitution
    for i in 1,...,N-1:
        b[i] = (b[i] - u[i] * b[p[i]]) / d[i]

    return b
```

# 3 Parallel GPU Implementation

Generally, neuroscientists are is interested in simulating a network of neurons. Therefore in each time step a large number of Hines matrices must be solved, since each neuron has its own dendritic tree. Because each Hines matrix can be solved independently, the existing solver in Arbor assigns one GPU thread to each matrix. However, especially for smaller networks or networks with very different kind of neurons, this might not expose enough parallelism and might lead to highly divergent branches.

In this section, we explain the new solver based on a finer level of parallelization, which leads to much better performance for smaller workloads and allows for easier work balancing of the Cuda blocks in general.

To get a finer level of parallelization, we also have to parallelize the solution of each matrix. The central idea is that the dependencies of the submatrices during the backward and forward substitution are not arbitrary but form a dependency tree. Each node in this tree represents a submatrix for which we perform a backward or forward substitution, respectively. An example of such a tree can be seen in Figure 1.1.

Using this dependency tree we can now easily see which computational tasks depend on results from other tasks and which tasks can be run independently in parallel. For the backward substitution we start at the lowest level of the tree and perform the backward substitution on the submatrices corresponding to the leaf nodes in parallel. Once all the intermediate results are computed, and thus all dependencies resolved, we can continue with the next higher level and perform the backward substitution on the next submatrices. This process is repeated until we arrive at the root node of the tree. Then we can solve the submatrix corresponding to the root node. Afterwards we start with the forward substitution on the children of the root node in a similar way we computed the backward substitution with the difference that we now move from top to bottom.

While we have to use synchronization methods to make sure that the computations on the previous level are done before we start with the next level, we can compute all dendrite branches on a level in parallel. Hence, the new solver does not assign an entire matrix to a thread, but instead each thread computes a backward or forward substitution on one dendrite branch. This means that each node in the dependency graph on one level is computed by exactly one thread.

When we parallelize the algorithm we have to take care of how to distribute the work to individual Cuda blocks. While different matrices can be solved independently from each other, we have to synchronize the computations between the levels of each matrix. In order to avoid global synchronization, we have to ensure that all computations related to the same matrix takes place in only one Cuda block so that block wide synchronization can be used instead.

# 4 Distribution of Branches into Blocks

In the previous section we discussed how we can parallelize the computation of each matrix using multiple threads. We now shortly discuss how the matrices get assigned to Cuda blocks so that they can be efficiently computed during the simulation. During this process it is important that all computations related to one matrix are computed by one single Cuda block so that the different backward and forward substitution steps can be synchronized on a per block basis.

As the number of neurons and their dendrites do not change over the simulation time we only have to compute the distribution of the matrices once during the setup phase. For a fixed block size we want to fit as many matrices into one block as possible to get a good workload for each Cuda block. Currently this is done in a "greedy" way which takes each matrix, counts how many dendrite branches there are on each level of the dependency tree and checks if they fit into a block. Since each GPU thread performs computations on one dendritic branch, we assign matrices to a Cuda block as long as each level has fewer dendritic branches than there are GPU threads in a block. When a matrix doesn't fit into a Cuda block any longer, we start a new Cuda block until all matrices have been distributed.

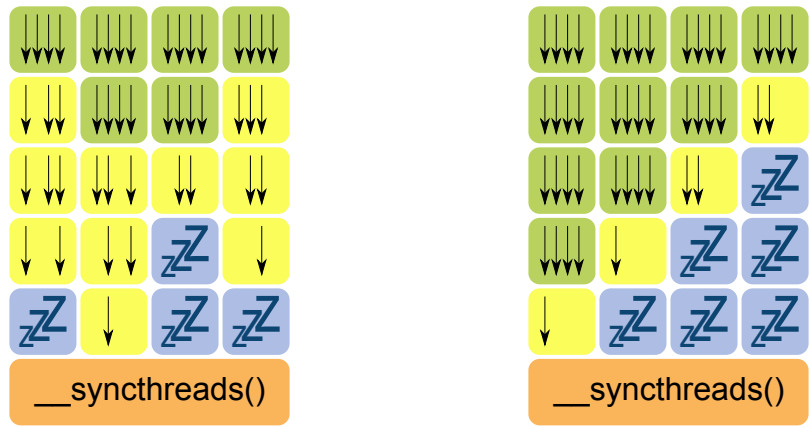

Figure 4.1: To reduce the branch divergence in the threads one can sort the submatrices on each level of the dependency tree by their size. Here we see one block with 4 warps with 4 threads each. Every thread has to perform up to five elimination steps. On the left the 16 submatrices are distributed randomly. We can see that many blocks are kept busy even though only few threads have to do actual work, leading to bad recourse utilization. On the right we sorted the submatrices by their size. This lead to less divergence and better recourse utilization.

Since each dendrite branch on a level of the dependency tree is computed in parallel we can choose which thread computes which branch. We can thus optimize to reduce thread divergence. During each substitution step, each thread executes a `for` loop where the upper bound depends on the size of the submatrix. The size corresponds to the number of compartments in which we subdivided the corresponding dendrite branch. If the size of the submatrices varies between different threads they divergence which hurts performance because idle threads waste resources without doing any useful work. As a first step to reduce the divergence we sort all submatrices on a level in a block by their size. Thus it becomes more likely that submatrices in a warp have a similar size. The effect is illustrated in Figure 4.1.

## 5 Memory Layout and Memory Access

To get good memory access patterns in the `for` loops of the solver, we store the data for each branch in an interleaved format so that in each step of the solver we can load each matrix entry needed for a substitution step with one continuous memory access for the entire warp. For each level in the dependency trees we store an offset into the buffer where the data for the branches start. Starting at the offset, each branch stores its data with a stride such that each memory read in a wrap is continuous. The stride is equal to the number of branches on the current level. Since some branches are shorter than others, this introduces some unused memory between the data for the longer branches but simplifies the memory access pattern because the stride remains constant. Also due to the branch cutting optimization, we expect most branches to be of similar size, which means that the additional memory needed should not be much more compared to a compact storage format.

While the Hines algorithm presented in Listing 2 works, it has the drawback that it needs many reads from the parent array $p$. To reduce the number of read accesses, we can make use of the fact that each branch is represented by a tridiagonal submatrix. We can perform the backward and forward substitution on each branch without reading from $p$ since we know that the parent of compartment $i$ is given implicitly by $i - 1$ as it was in the Thomas algorithm. To eliminate the off-diagonal coupling entries at the end of each substitution step we only have to read from the parent array. Thus, it is sufficient to store only one parent compartment for each branch.

## 6 Work Balancing

While the fine solver already shows a major speedup for smaller workloads, the profiler reports that a large percentage of branches diverge and that most warps are stalled due to synchronization barriers. One reason for this is that the length of the dendritic branches on one level can vary a lot. This means that in each block there are threads that have considerably more work to do and are stalling all other threads that are waiting on the next synchronization barrier. This stalling can also be seen in Figure 4.1, where many threads can not continue their work because they have to wait for one thread which still has to do more steps to finish it's task.

To overcome this problem, we introduce an optimization pass during the work distribution step before we assign the matrices to the Cuda blocks. We use two strategies to optimize the work distribution:

> *Balance the dependency tree* so that the depth of the tree is minimized. This reduces the number of synchronization points in the solver for badly balanced trees and leads to more tasks that can be computed in parallel since there are more nodes per level.

> *Split long branches* to balance the work in the Cuda blocks, such that the work distribution becomes more uniform and less threads are stalled on few computationally more expensive tasks.

To balance the dependency tree, we take each neuron and create a tree where each compartment is connected to its neighboring compartments as we already have done in Figure 1.1. To balance the tree, we have to find a new root node that minimizes the depth of the tree and permute all unknowns such that they form a new Hines matrix based on the new root node. In the example, this would be the compartment with index 7. To get a new Hines matrix, we have to renumber the compartments based on this new root node. To preserve the tridiagonal submatrices, we have to ensure that compartments in a dendrite branch get continuous indices. This can be achieved by numbering all branches breadth first and then sorting all compartments by branch index and distance to the root.

To improve the work balancing in each Cuda block, we split branches that are too long. Thus computationally expensive tasks in the dependency tree are split so that we can distribute the work over multiple levels of the dependency tree. Instead of one long branch, we now model it as two shorter branches that are attached to each other. This improves the balancing and allows idle thread to continue with their computations earlier at the cost of inserting one synchronization inside one branch.

Currently the branch splitting is performed on a per level basis. On each level we compute the average length of all dendritic branches on this level and cut all branches that are longer than average. Then we proceed to the next level and repeat the splitting until we reach the lowest level of the tree. The computation is done on a per level basis since cutting a branch at a specific level causes all child branches to move one level down. This means that we don't know which level a branch will be in until we finished cutting all parent branches.

Although these optimization steps lead to a changed tree structure, the branch distribution of the matrices as described before remains unchanged.

The complexity of the optimizations steps is linear in the number of neurons $N$. For the tree balancing this can be easily seen since this optimization works independently on each dendritic tree. For branch splitting, this can be seen by taking a look at the algorithm. On each level, we compute the average length of all branches on this level in all dendritic trees. For each level, we have to iterate over all trees. Even though the number of levels can increase during the algorithm due to branch splitting, the maximum number of levels is bounded by the number of compartments in each tree and thus is independent from the number of neurons. Because both optimizations steps are computed sequentially we get a total complexity of $\mathcal{O}(N)$.

## 7 BENCHMARKS AND RESULTS

To measure the performance of the new solver we compared the fine solver, with and without branch cutting, to the old solver that assigns one matrix per GPU thread. In the plots they will be referred to as "fine", "fine with opt" and "old". For the benchmarks we choose the ring network example in the arbor repository [2]. In this examples all dendritic trees are created with a specific number of levels. On each level branches are split into two child branches with a predefined probability. If a branch does not split it will just continue on the next level. This ensures that the different dendritic trees are randomized. The length of each branch decreases the further it is removed from the soma.

In the benchmarks, we have considered different settings with different branching probabilities, branch lengths and number of compartments per branch. For each setup we have varied the number of neurons in the network such that the performance of the solvers can be compared for both small and large networks.

Figure 7.1 shows a typical result from the benchmarks. For large numbers of neurons we can see the expected linear scaling of the solver: twice as many neurons means twice the number of matrices to solve. For smaller numbers of neurons, all methods scale better than linearly. Doubling the number of neurons leads to less than twice as much time spent in the solver. The reason for this is that for small numbers of neurons the GPU cannot be fully utilized. More neurons can therefore be computed in parallel without increasing the runtime by the same amount.

We also see that for small workloads the fine solver performs significantly better than the old solver. This is related to the finer level of parallelism in the fine solver compared to the old solver. As the fine solver has more but smaller computational tasks which can be computed in parallel, they can make use of more processing units of the GPU and utilize it more efficiently. In contrast the old solver has less larger computational tasks, which means that it might not be able to use all units on the GPU for small workloads.

This also explains why the fine solver picks up the linear scaling earlier than the old solver. Because the fine solver is able to fully utilize the GPU for smaller workloads, it starts the linear scaling regime with a smaller number of neurons, while the old solver still can't use all computational resources. Once the workload is large enough both methods scale linearly. For speedup, this means that we usually have a very large speedup for smaller workloads, which decreases to a constant level once both methods reach linear scaling.

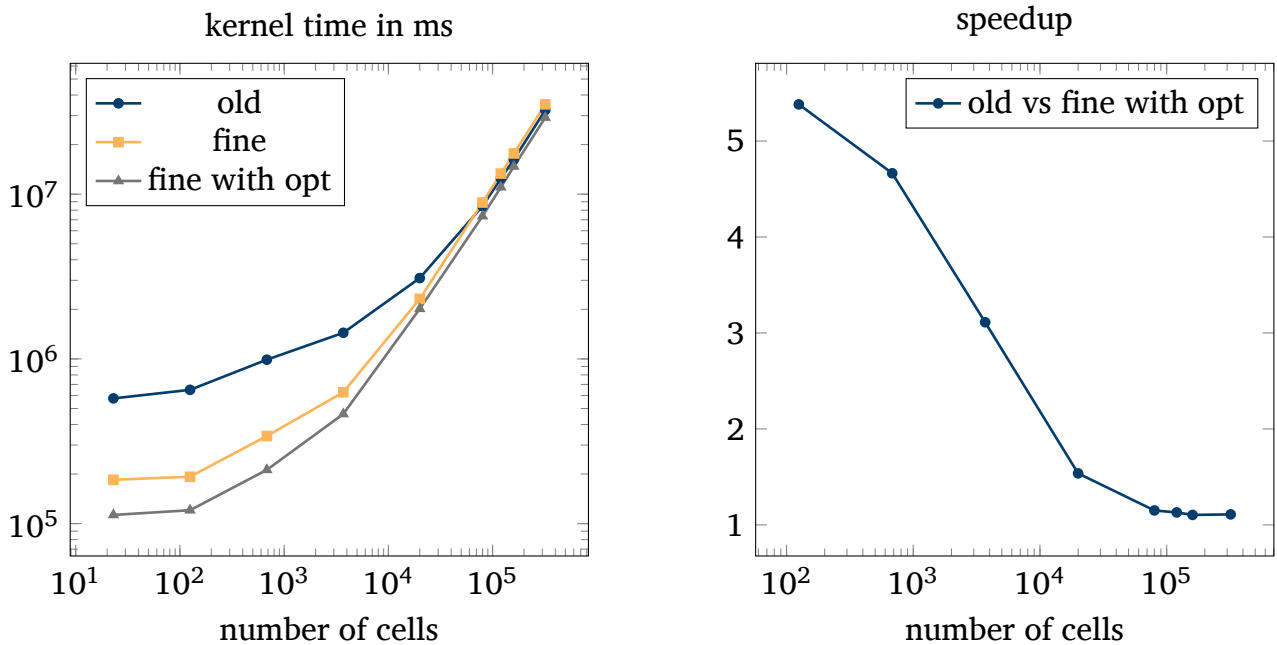

Figure 7.1: Typical benchmark result where the optimized fine solver only uses branch splitting. On the left we have the total time spent in kernels related to the solver for a different number of neurons. This includes the Hines solver kernel and the permutation of the data into from a solver specific format which allows for more efficient memory access. On the right we see the corresponding speedup over the old solver of the fine solver with optimization which reaches a constant level for large workloads.

Furthermore, the fine solver seems to have an advantage the more the dendritic trees of the neurons in a network differ from each other. Figure 7.2 shows a comparison of the speedup for different branching rates of the dendritic tree. If we lower the branching probability of the dendritic branches, the trees become more irregular and the fine solver often shows better performance than the old solver even for a larger number of neurons. A reason for this may be that the old solver suffers from high branch divergence when the length of the branches and number of branches on each level can vary a lot. The fine solver can more easily cope with this problem because the the branches on each level are computed in parallel and longer branches can be cut into smaller branches. Thus the shape of the trees is not as important for the fine solver.

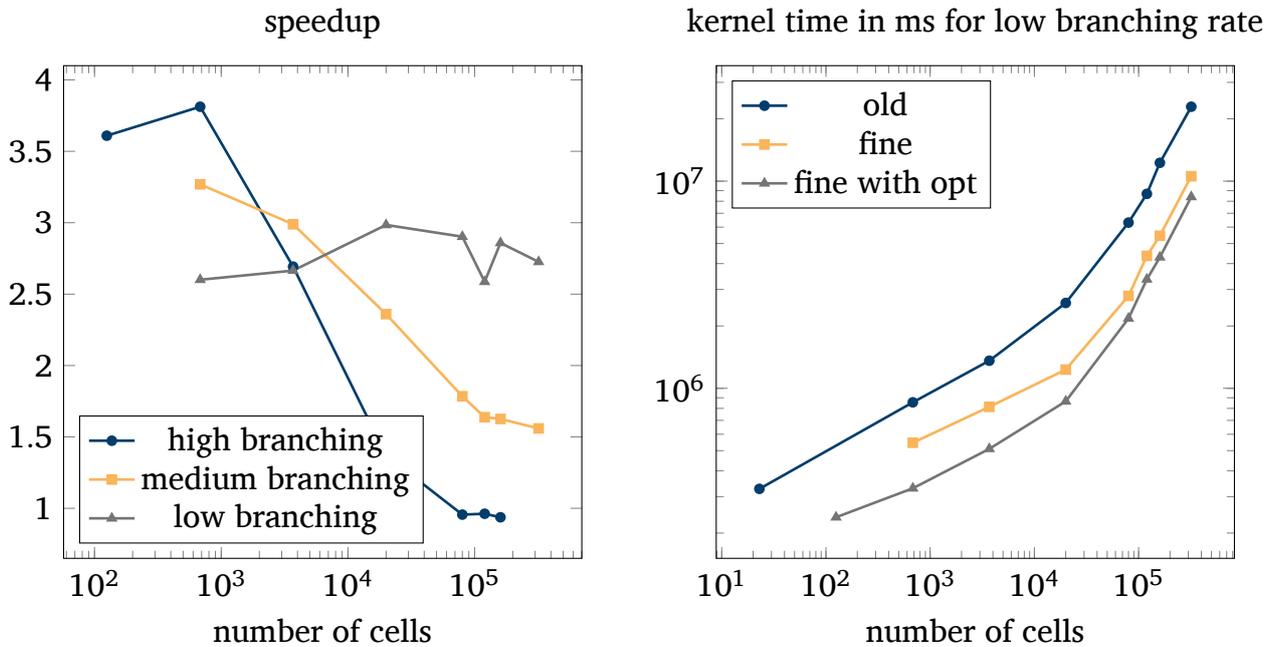

Figure 7.2: Speedup of the optimized fine solver over the old method for different branching settings. We can see that the speedup on the left is higher the lower the branching probability and thus the more irregular the dendritic trees become. On the right we see the time spent in the solver kernels for the low branching rate.

# 8 Conclusion and Outlook

The benchmarks clearly show that the new fine solver performs significantly better on small workloads due to its fine grained parallelization approach. For larger workloads it seams to perform similarly to the old solver or better, depending on how randomized the dendritic trees are. Furthermore it allows for simpler work balancing, which is especially interesting for long running simulations and networks where the neurons vary a lot.

More benchmarks are needed to compare the solvers on networks where the neurons have a more randomized structure to see how big the effect of work balancing is.

The current branch splitting is based on taking an overall average of the branch length on each level. In the future more sophisticated heuristics may be used to decide when to split a branch. Some intermediate packing of the matrices in a block can be used to compute more local heuristics, like per block averages.

While the fine solver can greatly improve the work balancing for each solver step, another balancing problem appears. As the number of cells in a Cuda block is currently limited by the sum of branches per level in the dependency tree, some levels might have much more work to do than others. One reason is that we currently pack all branches on one level of the matrices in the block into one computational block. Once one level reaches the maximum number of branches, a new block is started even though other levels might still have space left. One improvement would be to optimize the packaging in each block by better distributing work better between the computational steps such that the number of branches in each level gets closer to the maximum number of branches. One solution would be to allow subtrees in each dependency tree to move one level down if this improves the balancing in the different solver steps of the fine solver.

# 9 Acknowledgment

I want to thank my supervisors Alexander Peyser, Ben Cumming and Anne Küsters for making this project possible and the in depth discussion of all upcoming questions. Furthermore, I would like to thank Ivo Kabadschow and Laura Morgenstern for organizing the Guest Student Programme and Dominik Göddeke for suggesting this program.